\documentclass[runningheads,orivec]{llncs}

\usepackage[ruled,vlined,linesnumbered]{algorithm2e}
\usepackage{amsmath}
\usepackage{amssymb}
\usepackage{amsfonts}
\usepackage{cancel}
\usepackage{cite}
\usepackage{color}
\usepackage{comment}
\usepackage{graphicx}
\usepackage{hyperref}
\usepackage{subcaption}
\usepackage{tikz}
\usetikzlibrary{arrows,decorations,patterns,petri,positioning,shapes}
\usepackage{wasysym}
\usepackage{wrapfig}

\tikzset{
	silent/.style={
		minimum width = 2mm,
		minimum height = 5mm,
		fill = black
	}
}
\newlength{\hatchspread}
\newlength{\hatchthickness}
\newlength{\hatchshift}
\newcommand{\hatchcolor}{}
\tikzset{hatchspread/.code={\setlength{\hatchspread}{#1}},
	hatchthickness/.code={\setlength{\hatchthickness}{#1}},
	hatchshift/.code={\setlength{\hatchshift}{#1}},
	hatchcolor/.code={\renewcommand{\hatchcolor}{#1}}}
\tikzset{hatchspread=3pt,
	hatchthickness=0.4pt,
	hatchshift=0pt,
	hatchcolor=black}
\pgfdeclarepatternformonly[\hatchspread,\hatchthickness,\hatchshift,\hatchcolor]
{custom north west lines}
{\pgfqpoint{\dimexpr-2\hatchthickness}{\dimexpr-2\hatchthickness}}
{\pgfqpoint{\dimexpr\hatchspread+2\hatchthickness}{\dimexpr\hatchspread+2\hatchthickness}}
{\pgfqpoint{\dimexpr\hatchspread}{\dimexpr\hatchspread}}
{
	\pgfsetlinewidth{\hatchthickness}
	\pgfpathmoveto{\pgfqpoint{0pt}{\dimexpr\hatchspread+\hatchshift}}
	\pgfpathlineto{\pgfqpoint{\dimexpr\hatchspread+0.15pt+\hatchshift}{-0.15pt}}
	\ifdim \hatchshift > 0pt
	\pgfpathmoveto{\pgfqpoint{0pt}{\hatchshift}}
	\pgfpathlineto{\pgfqpoint{\dimexpr0.15pt+\hatchshift}{-0.15pt}}
	\fi
	\pgfsetstrokecolor{\hatchcolor}
	\pgfusepath{stroke}
}

\newcommand{\splitatcommas}[1]{%
  \begingroup
  \begingroup\lccode`~=`, \lowercase{\endgroup
    \edef~{\mathchar\the\mathcode`, \penalty0 \noexpand\hspace{0pt plus 1em}}%
  }\mathcode`,="8000 #1%
  \endgroup
}

\newcommand{\xrightarrowdbl}[2][]{%
  \xrightarrow[#1]{#2}\mathrel{\mkern-14mu}\rightarrow
}

\newcommand{\allTrails}{\ensuremath{\Gamma}}

\newcommand{\edges}{\ensuremath{E}}

\newcommand{\emptySequence}{\ensuremath{\epsilon}}
\newcommand{\graph}{\ensuremath{G}}
\newcommand{\labelFunc}{\ensuremath{\ell}}
\newcommand{\labelFuncPTWF}{\ensuremath{\tilde{\labelFunc}}}
\newcommand{\labels}{\ensuremath{\Sigma}}
\newcommand{\lang}{\ensuremath{\mathcal{L}}}
\newcommand{\langNet}{\ensuremath{\lang_{\univNets}}}
\newcommand{\langNetVis}{\ensuremath{\langNet^{\nu}}}
\newcommand{\langPT}{\ensuremath{\lang_{\univProcessTree}}}
\newcommand{\marking}{\ensuremath{M}}
\newcommand{\naturals}{\ensuremath{\mathbb{N}}}
\newcommand{\net}{\ensuremath{N}}
\newcommand{\netArcs}{\ensuremath{F}}

\newcommand{\place}{\ensuremath{p}}
\newcommand{\places}{\ensuremath{P}}

\newcommand{\pnptPattern}{\ensuremath{\theta}}
\newcommand{\pnptRedution}{\ensuremath{\Theta}}
\newcommand{\powerSet}{\ensuremath{\mathcal{P}}}
\newcommand{\processTree}{\ensuremath{Q}}
\newcommand{\processTreeOp}{\ensuremath{\oplus}}
\newcommand{\ptNet}{\ensuremath{\tilde{\net}}}

\newcommand{\ptWfNet}{\ensuremath{\tilde{\wfNet}}}
\newcommand{\ptToWfNetTranslator}{\ensuremath{\lambda}}
\newcommand{\ptToWfNetTranslatorTransB}{\ensuremath{\hat{\ptToWfNetTranslator}}}
\newcommand{\ptWfNetUnfolding}{\ensuremath{\Lambda}}
\newcommand{\reachable}{\ensuremath{\mathcal{R}}}
\newcommand{\silent}{\ensuremath{\tau}}
\newcommand{\sequence}{\ensuremath{\sigma}}
\newcommand{\systemNet}{\ensuremath{SN}}
\newcommand{\transition}{\ensuremath{t}}
\newcommand{\transitions}{\ensuremath{T}}
\newcommand{\univMultisets}{\ensuremath{\mathcal{M}}}
\newcommand{\univNets}{\ensuremath{\mathcal{N}}}
\newcommand{\univProcessTree}{\ensuremath{\mathcal{Q}}}
\newcommand{\univProcessTreeOperators}{\ensuremath{\bigoplus}}
\newcommand{\univPtNet}{\ensuremath{\univNets_{\univProcessTree}}}
\newcommand{\univPtSystemNet}{\ensuremath{\univSystemNet_{\univProcessTree}}}
\newcommand{\univPtWfNet}{\ensuremath{\univWfNets_{\univProcessTree}}}
\newcommand{\univSystemNet}{\ensuremath{\mathcal{SN}}}
\newcommand{\univWfNets}{\ensuremath{\mathcal{W}}}
\newcommand{\vertex}{\ensuremath{v}}
\newcommand{\vertices}{\ensuremath{V}}
\newcommand{\wfNet}{\ensuremath{W}}

\begin{document}
\title{Translating Workflow Nets to Process Trees:\\ An Algorithmic Approach}
\titlerunning{Translating Workflow Nets to Process Trees}
%
\author{Sebastiaan J. van Zelst\inst{1,2}}
\authorrunning{Sebastiaan J. van Zelst}
%
\institute{Fraunhofer Institute for Applied Information Technology (FIT), Germany\\ \and
RWTH Aachen University, Aachen, Germany \email{sebastiaan.van.zelst@fit.fraunhofer.de} 
}

\maketitle              
\begin{abstract}
Since their recent introduction, process trees have been frequently used as a process modeling formalism in many process mining algorithms. 
A process tree is a tree-based model of a process, in which internal vertices represent behavioral control-flow relations and leaves represent process activities.
A process tree is easily translated into a sound Workflow net (WF-net), however, the reverse is not the case. 
Yet, an algorithm that translates a WF-net into a process tree is of great interest, e.g., the explicit knowledge of the control-flow hierarchy in a WF-net allows one to more easily reason on its behavior.
Hence, in this paper, we present such an algorithm, i.e., it detects whether a WF-net corresponds to a process tree, and, if so, constructs it.
We prove that, if a process tree is discovered, the language of the process tree equals the language of the original WF-net.
Conducted experiments show, that the algorithm's corresponding implementation has a quadratic time-complexity in the size of the WF-net.
Furthermore, the experiments show strong evidence of process tree rediscoverability.
\keywords{Process Trees  \and Petri Nets \and Workflow nets \and Process Mining.}
\end{abstract}

\section{Introduction}
The research field of \emph{process mining}~\cite{DBLP:books/sp/Aalst16}, is concerned with distilling knowledge of the execution of processes, by analyzing the event data generated during the execution of these processes, i.e., stored in modern-day information systems.
In the field, different semi-automated techniques have been developed, that are able to distill processes knowledge, ranging from \emph{automated process discovery algorithms} to \emph{conformance checking algorithms}.
As processes are the cornerstone of process mining, so are the models that allow us to represent them, and/or reason on their behavior/quality.
Various process modeling formalisms exists, e.g., BPMN~\cite{DBLP:journals/infsof/DijkmanDO08}, EPCs~\cite{DBLP:journals/infsof/Aalst99}, etc., some of which are heavily used in industry.

Recently, process trees were introduced~\cite{DBLP:conf/simpda/AalstBD11}.
A process tree is a hierarchical representation of a process, corresponding to a rooted tree, i.e., a connected, undirected acyclic graph, with a designated root vertex.
The internal vertices represent how their children are related to each-other, in terms of control-flow behavior.
The leaves of the tree represent the activities of the process.
Consider \autoref{fig:example_tree} (\autopageref{fig:example_tree}), in which we depict an example process tree.
Its root vertex has label $\to$, specifying that first its left-most child, i.e., activity with $a$, needs to be executed, secondly its middle child, and, finally its right-most child.

Process trees are easily translated into other process modeling formalisms, e.g., \emph{Workflow nets (WF-nets)}.
By definition, a process tree corresponds to a \emph{sound WF-net}, i.e., a WF-net with desirable behavioral properties, e.g., the absence of deadlocks.
The reverse, i.e., given a WF-net, translating it into a process tree (if possible), is less trivial.
At the same time, obtaining such a translation is of interest, i.e., it allows us to discover control-flow aware hierarchical structures within a WF-net.
Such structures, can, for example, be used to hide certain parts of the model, i.e., leading to a more comprehensible view of the process model.
Furthermore, any algorithm optimized for process trees, e.g., by exploiting the hierarchical structure, is applicable to WF-nets of such a type.
For example, in~\cite{DBLP:journals/isci/LeeVMAS18}, it is shown that the computation time of \emph{alignments}~\cite{DBLP:journals/topnoc/ZelstBD18}, i.e., explanations of observed behavior in terms of a reference model, can be significantly reduced by applying Petri net decomposition, e.g., driven by model hierarchies.


In this paper, we present an algorithm that determines whether a given WF-net corresponds to a process tree, and, if so, constructs it.
We prove that, if a process tree is found, the original WF-net is sound, and, that the obtained process tree's language is equal to the language of the original WF-net.
A corresponding implementation, extending the process mining framework PM4Py~\cite{DBLP:journals/corr/abs-1905-06169}, is publicly available.
Using the implementation, we conducted several experiments, which show a quadratic time complexity in terms of the WF-net size.
Furthermore, our experiments indicate rediscoverability of process trees.

The remainder of this paper is structured as follows.
In~\autoref{sec:prelim}, we present preliminary concepts and notation.
In~\autoref{sec:method}, we present the proposed algorithm, including the proofs w.r.t soundness preservation and language preservation.
In~\autoref{sec:eval}, we evaluate our approach.
In~\autoref{sec:rel_work}, we discuss related work.
\autoref{sec:conclusion}, concludes the paper.

\section{Preliminaries}
\label{sec:prelim}
Given set $X$, $\powerSet(X){=}\left\{X'\subseteq{X}\right\}$ denotes its powerset.
Given a function $f{\colon}X{\to}Y$ and $X'{\subseteq}X$, we extend function application to sets, i.e., $f(X'){=}\{y\mid\exists{x{\in}X'}(f(x){=}y)\}$.
Furthermore, $f|_{X'}{\colon}X'{\to}Y$ restricts $f$ to $X'$.
A multiset over set $X$, i.e., $m{\colon}X{\to}\naturals{\cup}\{0\}$, contains multiple instances of an element.
We write a multiset as $m{=}[x_1^i,x_2^j,...,x_n^k]$, where $m(x_1){=}i,m(x_2){=}j,...,m(x_n){=}k$, for $i,j,...,k{>}1$ (in case, $m(x_i){=}1$, we omit its superscript; in case $m(x_i){=}0$, we omit $x_i$).
The set of all multisets over $X$ is written as $\univMultisets(X)$.
The sum  of two multisets $m_1,m_2$ is written as $m_1{\uplus}m_2$, e.g., $[x^2,y]{\uplus}[x^3,y,z]{=}[x^5,y^2,z]$, their difference is written as $m_1{\setminus}m_2$, e.g., $[x^2,y]{\setminus}[x,y,z]{=}[x]$.
A set is considered a multiset with each element appearing only once, hence, $\{x,y,z\}{\uplus}[x^2]{=}[x^3,y,z]$. 
$X^*$ denotes the set of all sequences over $X$.
We write $|\sequence|$ to denote the length of $\sequence{\in}X^*$, and, we write $\splitatcommas{\sequence{=}\langle\sequence(1),\sequence(2),...,\sequence(|\sequence|)\rangle}$, where $\sequence(i)$ denotes the element at position $i$ ($1{\leq}i{\leq}|\sequence|$).
$\emptySequence$ denotes the empty sequence, i.e., $|\emptySequence|{=}0$.
We extend the notion of element inclusion to sequences, e.g., $x{\in}\langle x,y,z\rangle$. 
Concatenation of sequences $\sequence,\sequence'{\in}X^*$ is written as $\sequence \cdot \sequence'$.
We let $\sequence{\leftrightharpoons}\sequence'$ denote the set of all possible order-preserving merges, i.e., the \emph{shuffle operator}, of $\sequence$ and $\sequence'$, e.g., given $\sequence_1{=}\langle b,p\rangle$, $\sequence_2{=}\langle m \rangle$, then $\splitatcommas{\sequence_1{\leftrightharpoons}\sequence_2{=}\left\{\langle b,p,m \rangle, \langle b,m,p \rangle, \langle m,b,p\rangle\right\}}$.
It is easy to see that $\sequence{\leftrightharpoons}\sequence'{=}\sequence'{\leftrightharpoons}\sequence$ (commutative).
We extend the shuffle operator to sets (and overload notation), i.e., given $S,S'{\in}X^*$, $S{\leftrightharpoons}S'{=}\{\sequence{\in}\sequence_1{\leftrightharpoons}\sequence_2\mid \sequence_1{\in}S_1{\wedge}\sequence_2{\in}S_2\}$.
Note that, $(\sequence{\leftrightharpoons}\sequence'){\leftrightharpoons}\{\sequence''\} = \{\sequence\}{\leftrightharpoons}(\sequence'{\leftrightharpoons}\sequence'')$ (associative), hence, we write the application of the shuffle operation on $n$ sequences as $\sequence_1{\leftrightharpoons}\sequence_2{\leftrightharpoons}{\cdots}{\leftrightharpoons}{\sequence_n}$. 
Similarly, we wirte $S_1{\leftrightharpoons}S_2{\leftrightharpoons}{\cdots}{\leftrightharpoons}S_n$ for sets of sequences $S_1,S_2,...S_n{\in}X^*$.
Given a function $f{\colon}X{\to}Y$ and a sequence $\sequence{\in}X^*$, we overload notation for function application, i.e., $\splitatcommas{f(\sequence){=}\langle(f(\sequence(1)),f(\sequence(2)), ..., f(\sequence{|\sequence|})\rangle}$.
Note that, this extends to sets of sequences, i.e., given $f{\colon}X\to Y$ and $X'{\subseteq}X^*$, $f(X'){=}\{\sequence{\in}Y^* \mid \exists{\sequence'{\in}X'}\left(f(\sequence'){=}\sequence\right)\}$.
Furthermore, given $X'{\subseteq}X$ and a sequence $\sequence{\in}X^*$, we define $\sequence_{\downarrow_{X'}}$, where (recursively) $\emptySequence_{\downarrow_{X'}}{=}\emptySequence$ and $(\langle x \rangle \cdot \sequence)_{\downarrow_{X'}}{=}x\cdot\sequence_{\downarrow_{X'}}$ if $x{\in}X'$ and $(\langle x \rangle \cdot \sequence)_{\downarrow_{X'}}{=}\sequence_{\downarrow_{X'}}$ if $x{\notin}X'$.

\subsection{Workflow Nets}
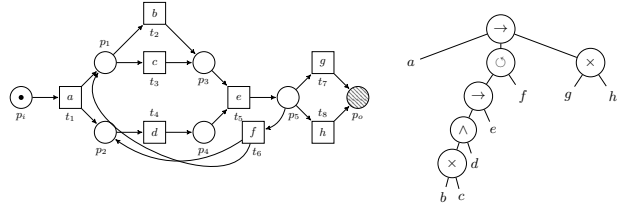
\begin{figure}[tb]
	\centering
	\begin{subfigure}[t]{0.4\textwidth}
	    \centering
    	\resizebox{\textwidth}{!}{
    		\begin{tikzpicture}
    		[node distance=1.35cm,
    		on grid,>=stealth',
    		bend angle=30,
    		auto,
    		every place/.style= {minimum size=6mm},
    		every transition/.style = {minimum size = 6mm}
    		]
    		
    		\node [place, tokens = 1] (start) [label=below:$\place_i$]{};
    		
    		\node [transition] (t1) [label=below:$\transition_1$, label=center:$a$, right = of start] {}
    		edge [pre] node[auto] {} (start);
    		
    		\node [place] (p1) [label=above:$\place_1$, above right = of t1] {}
    		edge [pre] node[auto] {} (t1);
    		
    		\node [transition] (t3) [label=below:$\transition_3$, label=center:$c$, right = of p1] {}
    		edge [pre] node[auto] {} (p1);
    		
    		\node [place] (p2) [label=below:$\place_2\ \ $, below right = of t1] {}
    		edge [pre] node[auto] {} (t1);	
    		
    		\node [transition] (t4) [label=above:$\transition_4$, label=center:$d$, right = of p2] {}
    		edge [pre] node[auto] {} (p2);
    		
    		\node [place] (p3) [label=below:$\place_3$, right = of t3] {}
    		edge [pre] node[auto] {} (t3);
    		
    		\node [place] (p4) [label=below:$\place_4$, right = of t4] {}
    		edge [pre] node[auto] {} (t4);
    		
    		\node [transition] (t5) [label=below:$\transition_5$, label=center:$e$, above right = of p4] {}
    		edge [pre] node[auto] {} (p3)
    		edge [pre] node[auto] {} (p4);
    		
    		\node [transition] (t2) [label=below:$\transition_2$, label=center:$b$, above = of t3] {}
    		edge [pre] node[auto] {} (p1)
    		edge [post] node[auto] {} (p3);	
    		
    		\node [place] (p5) [label=below:$\ \ \ \place_5$, right = of t5] {}
    		edge [pre] node[auto] {} (t5);
    		
    		\node [transition] (t6) [label=below:$\ \ \transition_6$, label=center:$f$, right = of p4] {}
    		edge [pre, bend right] node[auto] {} (p5)
    		edge [post, bend left = 35] node[auto] {} (p2)
    		edge [post, bend left = 100] node[auto] {} (p1);
    		
    		\node [transition] (t7) [label=below:$\transition_7$, label=center:$g$, above right = of p5] {}
    		edge [pre] node[auto] {} (p5);
    		
    		\node [transition] (t8) [label=above:$\transition_8$, label=center:$h$, below right = of p5] {}
    		edge [pre] node[auto] {} (p5);
    		
    		\node [place] (end) [label=below:$\place_o$, below right = of t7, pattern=custom north west lines,hatchspread=1.5pt,hatchthickness=0.3pt,hatchcolor=gray,] {}
    		edge [pre] node[auto] {} (t7)
    		edge [pre] node[auto] {} (t8);		
    		\end{tikzpicture}
    	}
    	\caption{WF-net $\wfNet_1$~\cite{DBLP:books/sp/Aalst16} with initial marking $[\place_i]$ and final marking $[\place_o]$. } 
	    \label{fig:example_petri_net}
	\end{subfigure}\quad
	\begin{subfigure}[t]{0.25\textwidth}
    \centering
    \resizebox{\textwidth}{!}{
    \begin{tikzpicture}[level/.style={sibling distance=20mm/#1, level distance = 7.5mm}]
      \node [circle, draw] (l0_seq){$\to$}
  child {node [] (a) {$a$}}
  child {node [circle,draw] (l1_loop) {$\circlearrowleft$}
    child {node [circle,draw] (l2_seq) {$\to$}
      child {node [circle,draw] (l3_par) {$\wedge$}
        child {node [circle,draw] (l4_xor) {$\times$}
          child{node [] (b) {$b$}}
          child{node [] (c) {$c$}}
        }
        child {node [] (d) {$d$}}
      } 
      child {node [] (e) {$e$}}
    }
    child {node [] (f) {$f$}}
  }
  child {node [circle,draw] (l1_xor) {$\times$}
    child{node [] (g) {$g$}}
    child{node [] (h) {$h$}}
  };
\end{tikzpicture}
}
    \caption{Process tree~\cite{DBLP:books/sp/Aalst16}, describing the same language as the WF-net in \autoref{fig:example_petri_net}. }
    \label{fig:example_tree}
\end{subfigure}
\caption{Two process models describing the same language, i.e., a WF-net (\autoref{fig:example_petri_net}), and, a process tree (\autoref{fig:example_tree}).}
\label{fig:example_models}
\end{figure}

Workflow nets (WF-nets)~\cite{DBLP:journals/jcsc/Aalst98} extend the more general notion of Petri nets~\cite{murata1989petri}.
A Petri net is a directed bipartite graph containing two types of vertices, i.e., places and transitions.
Places are visualized as circles, transitions are visualized as boxes.
Places only connect to transitions, and vise-versa.
Consider \autoref{fig:example_petri_net}, depicting an example Petri net (which is also a WF-net).
We let $\net{=}(\places,\transitions,\netArcs,\labelFunc)$ denote a (labelled) Petri net, where, $\places$ denotes a set of places, $\transitions$ denotes a set of transitions and $\netArcs{\subseteq}(\places{\times}\transitions)\cup(\transitions{\times}\places)$ represents the arcs.
Furthermore, given a set of labels $\labels$ and symbol $\silent{\notin}\labels$, $\labelFunc{\colon}\transitions{\to}\labels{\cup}\{\tau\}$ is the net's labelling function, e.g., in \autoref{fig:example_petri_net}, $\labelFunc(\transition_1){=}a$, $\labelFunc(\transition_2){=b}$, etc.
Given an element $x{\in}\places{\cup}\transitions$, ${\bullet}x=\left\{y \mid (y,x){\in}\netArcs\right\}$ denotes the \emph{pre-set} of $x$, whereas $x{\bullet}{=}\left\{y| (x,y){\in}\netArcs\right\}$ denotes its \emph{post-set}, e.g., in \autoref{fig:example_petri_net}, ${\bullet}{\transition_1}{=}\left\{\place_i\right\}$ and $\place_1{\bullet}{=}\left\{\transition_2,\transition_3\right\}$.
We lift the $\bullet$-notation to the level of sets, i.e., given $X{\subseteq}\places{\cup}\transitions$, $\bullet{X}{=}\left\{y \mid \exists{x{\in}X}\left(y{\in}{\bullet}x \right) \right\}$ and ${X}{\bullet}{=}\left\{y \mid \exists{x{\in}X}\left(y{\in}x{\bullet} \right) \right\}$.
Let $\net{=}(\places,\transitions,\netArcs,\labelFunc)$, let $\places'{\subseteq}\places$, $\transitions'{\subseteq}\transitions$ and $\netArcs'{\subseteq}\netArcs$.
The net $\net'{=}\left(\places',\transitions',\netArcs',\labelFunc|_{\transitions'}\right)$ is a subnet of $\net$, written $\net'{\sqsubseteq}\net$.
We let $\univNets$ denote the universe of Petri nets.

The \emph{state} of a Petri net, is expressed by means of a \emph{marking}, i.e., a multiset of places.
A marking is visualized by drawing the corresponding number of dots in the place(s) of the marking, e.g., the marking in \autoref{fig:example_petri_net} is $[\place_i]$ (one black dot is drawn inside place $\place_i$).
Given a Petri net $\net=(\places,\transitions,\netArcs,\labelFunc)$ and marking $\marking{\in}\univMultisets(\places)$, $(\net,\marking)$ denotes a \emph{marked net}.
Given a marked net $(\net,\marking)$, a transition $\transition{\in}\transitions$ is \emph{enabled}, written $(\net,\marking){[}{\transition}{\rangle}$, if $\forall{\place{\in}{\bullet}{\transition}}\left(\marking(\place){>}0\right)$.
An enabled transition can \emph{fire}, leading to a new marking $\marking'{=}\left(\marking{\setminus}{\bullet}\transition\right){\uplus}{\transition}{\bullet}$, written $(\net,\marking){\xrightarrow{\transition}}(\net,\marking')$.
A sequence of transition firings $\sequence{=}\langle\transition_1,\transition_2,...,\transition_n\rangle$ is a \emph{firing sequence} of $(\net,\marking)$, yielding marking $\marking'$, written $(\net,\marking){\xrightarrowdbl{\sequence}}(\net,\marking')$, iff $\exists\marking_1,\marking_2,...,\marking_{n-1}{\in}\univMultisets(\places)$ s.t. $\splitatcommas{(\net,\marking){\xrightarrow{\transition_1}}(\net,\marking_1){\xrightarrow{\transition_2}}(\net,\marking_2){\cdots}(\net,\marking_{n-1}){\xrightarrow{\transition_n}}(\net,\marking')}$.
We write $(\net,\marking){\rightsquigarrow}(\net,\marking')$, in case $\exists{\sequence{\in}\transitions^*}\left((\net,\marking){\xrightarrowdbl{\sequence}}(\net,\marking')\right)$.
$\splitatcommas{\langNet(\net,\marking,\marking') = \left\{\sequence{\in}\transitions^* \mid (\net,\marking){\xrightarrowdbl{\sequence}}(\net,\marking')\right\}}$ denotes all firing sequences starting from marking $\marking$, leading to marking $\marking'$.
The \emph{labelled-language} of $\net$, conditional to markings $\marking,\marking'$, is defined as $\splitatcommas{\langNetVis(\net,\marking,\marking'){=}\labelFunc(\langNet(\net,\marking,\marking'))_{\downarrow_{\labels}}}$.
$\splitatcommas{\reachable(\net,\marking) = \left\{\marking'{\in}\univMultisets(\places) \mid \exists{\sequence{\in}\transitions^*}\left((\net,\marking)\xrightarrowdbl{\sequence}(\net,\marking')\right)\right\}}$ denotes the reachable markings.

Given a Petri net $\net{=}\left(\places,\transitions,\netArcs,\labelFunc\right)$, and designated initial and final marking $\splitatcommas{\marking_i,\marking_f{\in}\univMultisets(\places)}$, triple $\systemNet{=}(\net,\marking_i,\marking_f)$ denotes a \emph{system net}.
As system net $\systemNet{=}(\net,\marking_i,\marking_f)$ is formed by $\net$, we write $\systemNet$ as a replacement for $\net$, e.g., $(\systemNet,\marking)$ denotes a marked system net.
Clearly, $\reachable(\systemNet,\marking)$, $\langNet(\systemNet,\marking,\marking')$, etc., are readily defined. 
$\univSystemNet$ denotes the universe of system nets.

A WF-net is a special type of Petri net, i.e., it has one unique start and one unique end {place}.
Furthermore, every place/transition in the net is on a path from the start to the end place.
\begin{definition}[Workflow net (WF-net)]
\label{def:wf_net}
Let $\net=(\places,\transitions,\netArcs,\labelFunc){\in}\univNets$ and let $\place_i{\neq}\place_o{\in}\places$. 
Tuple $\wfNet{=}(\places,\transitions,\netArcs,\place_i,\place_o,\labelFunc)$ is a Workflow net (WF-net), iff:
\begin{enumerate}
    \item ${\bullet}\place_i{=}\emptyset{\wedge}\nexists{\place{\in}\places{\setminus}\{\place_i\}}\left({\bullet}\place{=}\emptyset\right)$; $\place_i$ is the unique source place.
    \item $\place_o{\bullet}{=}\emptyset{\wedge}\nexists{\place{\in}\places{\setminus}\{\place_o\}}\left(\place{\bullet}{=}\emptyset\right)$; $\place_o$ is the unique sink place.
    \item Each element in $\places{\cup}\transitions$ is on a path from $\place_i$ to $\place_o$.
\end{enumerate}
We let $\univWfNets{\subset}\univNets$ denote the universe of WF-nets.
\end{definition}

Of particular interest are \emph{sound WF-nets}, i.e., WF-nets that are, by definition, guaranteed to be free of \emph{deadlocks} and \emph{livelocks}.
We formalize the notion of \emph{soundness} in \autoref{def:wf_sound}.
\begin{definition}[Soundness]
\label{def:wf_sound}
Let $\wfNet{=}(\places,\transitions,\netArcs,\place_i,\place_o,\labelFunc){\in}\univWfNets$.
$\wfNet$ is \emph{sound} iff:
\begin{enumerate}
    \item $(\wfNet,[\place_i])$ is \emph{safe}, i.e., $\forall{\marking{\in}\reachable(\wfNet,[\place_i])}\left(\forall{\place{\in}\places}\left(\marking(\place){\leq}1\right)\right)$,
    \item \label{req:reach_final} $[\place_o]$ can always be reached, i.e., $\forall{\marking{\in}\reachable(\wfNet,[\place_i]})\left((\wfNet,\marking){\rightsquigarrow}(\wfNet,[\place_o])\right)$. 
    \item Each $\transition{\in}\transitions$ is enabled, at some point, i.e., $\forall{\transition{\in}\transitions}\left(\exists{\marking{\in}\reachable(\wfNet,[\place_i])}\left(\marking[\transition\rangle\right)\right)$.
\end{enumerate}
\end{definition}

Observe that, the Petri net depicted in \autoref{fig:example_petri_net}, is a sound WF-net. 

\subsection{Process Trees}
Process trees allow us to model processes, that comprise a control-flow hierarchy.
A process tree is a mathematical tree, where the internal vertices are \emph{operators} and leaves are (non-observable) \emph{activities}.
Operators specify how their children, i.e., sub-trees, need to be combined, from a control-flow perspective.
\begin{definition}[Process Tree]
\label{def:process_tree}
Let $\labels$ denote the {universe of (activity) labels} and let $\silent{\notin}\labels$.
Let $\univProcessTreeOperators$ denote the universe of \emph{process tree operators}.
A process tree $\processTree$, is defined (recursively) as any of:
\begin{enumerate}
    \item $x{\in}\labels{\cup}\{\silent\}$; an (non-observable) activity,
    \item ${\processTreeOp}(\processTree_1,...,\processTree_n)$, for $\processTreeOp{\in}\univProcessTreeOperators$, $n{\geq}1$, where $\processTree_1,...,\processTree_n$ are process trees;
\end{enumerate}
We let $\univProcessTree$ denote the universe of process trees.
\end{definition}

Several operators (elements of $\univProcessTreeOperators$) can be defined, however, in this work, we focus on four basic operators, i.e., the $\to$, $\times$, $\wedge$ and $\circlearrowleft$-operator.
The \emph{sequence} operator ($\to$) specifies sequential behavior.
First its left-most child is executed, then its second left-most child, ..., and finally its right-most child.
For example, the root operator in \autoref{fig:example_tree}, specifies that first activity $a$ is executed, then its second sub-tree ($\circlearrowleft$) and then its third sub-tree ($\times$).
The \emph{exclusive choice} operator $(\times)$, specifies an exclusive choice, i.e., one (and exactly one) of its sub-trees is executed.
Concurrent/parallel behavior is represented by the \emph{concurrency operator} ($\wedge$), i.e., all children are executed simultaneously/in any order. 
Repeated behavior is represented by the \emph{loop operator} $\circlearrowleft$.
The $\to$, $\times$ and $\wedge$-operator have an arbitrary number of children.
The $\circlearrowleft$-operator has two children.
Its left child (the ``do-child'') is \emph{always} executed.
Secondly, executing its right child (the ``re-do-child'') is optional.
After executing the re-do-child, we again execute the do-child.
We are allowed to repeat this, yet, we always finish with the do-child.

Given a process tree $\processTree{\in}\univProcessTree$, its language is of the form $\langPT(\processTree){\subseteq}\labels^*$. 
\begin{definition}[Process Tree Language]
\label{def:pt_lang}
Let $\processTree{\in}\univProcessTree$ be a process tree.
The language of $\processTree$, i.e., $\langPT(\processTree){\subseteq}\labels^*$, is defined recursively as:
\begin{enumerate}
    \item $\langPT(\processTree){=}\{\emptySequence\}$, if $\processTree{=}\silent$,
    \item $\langPT(\processTree){=}\{ \langle a \rangle \}$, if $\processTree{=}a{\in}\labels$,
    \item $\langPT(\processTree){=}\{ \sequence{=}\sequence_1{\cdot}\sequence_2{\cdots}\sequence_n \mid \sequence_1{\in}\langPT(\processTree_1), \sequence_2{\in}\langPT(\processTree_2), ..., \sequence_n{\in}\langPT(\processTree_n) \}$, if $\processTree = {\to}\left(\processTree_1,\processTree_2,...,\processTree_n\right)$,
    \item $\langPT(\processTree){=}\bigcup\limits_{i=1}^n\langPT(\processTree_n)$, if $\processTree{=}{\times}\left(\processTree_1,\processTree_2,...,\processTree_n\right)$
    \item $\langPT(\processTree){=}\langPT(\processTree_1)\leftrightharpoons\langPT(\processTree_2)\cdots\langPT(\processTree_n)$, if $\processTree{=}{\wedge}\left(\processTree_1,\processTree_2,...,\processTree_n\right)$
    \item $\langPT(\processTree){=}\{\sequence_1 \cdot \sequence'_1 \cdot \sequence_2 \cdot \sequence'_2 \cdots \sequence_n \mid n \geq 1 \wedge \forall{1{\leq}i{\leq}m}\left(\sequence_i{\in}\langPT(\processTree_1)\right) \wedge \forall{1{\leq}i{<}m}\left(\sequence'_i{\in}\langPT(\processTree_2)\right) \}$, if $\processTree{=}{\circlearrowleft}\left(\processTree_1,\processTree_2\right)$.
\end{enumerate}
\end{definition}

\begin{figure}[tb]
    \centering
    \includegraphics[width=0.85\textwidth]{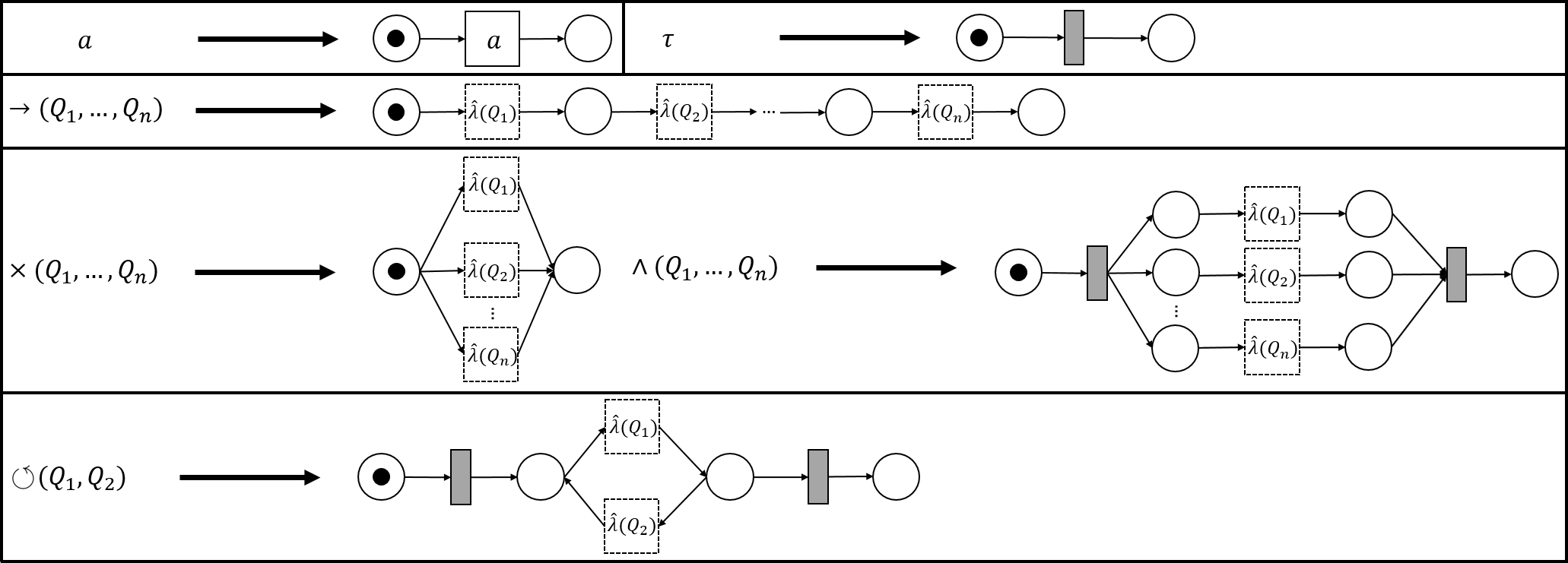}
    \caption{Example instantiations of $\ptToWfNetTranslator$ (\autoref{def:pt_pn_transformation}). The $\ptToWfNetTranslator$-functions for operators are defined recursively, using the $\ptToWfNetTranslatorTransB$-values of their children, i.e., a place ``entering''/``exiting'' a $\ptToWfNetTranslatorTransB(\processTree_i)$ fragment, connects to $\place_i{\bullet}$/${\bullet}{\place_o}$ (respectively) of $\ptToWfNetTranslator(\processTree_i)$.}
    \label{fig:pt_pn_translation}
\end{figure}

The process tree operators considered ($\to$, $\times$, $\wedge$ and $\circlearrowleft$), are easily translated to sound WF-nets, cf. \autoref{fig:pt_pn_translation}.
Hence, we define a generic process tree to WF-net translation function, s.t., the language of the two is the same.
\begin{definition}[Process Tree Transformation Function]
\label{def:pt_pn_transformation}
Let $\processTree{\in}\univProcessTree$ be a process tree.
A process tree transformation function $\ptToWfNetTranslator$, is a function $\ptToWfNetTranslator{\colon}\univProcessTree{\to}\univWfNets$, s.t., $\langNetVis(\ptToWfNetTranslator(\processTree)){=}\langPT(\processTree)$.
We let  $\ptToWfNetTranslatorTransB{\colon}\univProcessTree{\to}\univNets$, where, given $\ptToWfNetTranslator(\processTree){=}\wfNet{=}(\places,\transitions,\netArcs,\place_i,\place_o,\labelFunc)$, $\ptToWfNetTranslatorTransB(\processTree){=}(\places',\transitions,\netArcs',\labelFunc)$, with, $\places'{=}\place{\setminus}\{\place_i,\place_o\}$ and $\netArcs'{=}\netArcs{\setminus}\left(\{(\place_i,\transition){\in}\netArcs\}{\cup}\{(\transition,\place_o){\in}\netArcs\}\right)$.
\end{definition}

Given an arbitrary process tree $\processTree{\in}\univProcessTree$, there are several ways to translate it to a (sound) WF-net $\wfNet$, s.t., $\langNetVis(\wfNet){=}\langPT(\processTree)$, i.e., instantiating $\ptToWfNetTranslator$ and $\ptToWfNetTranslatorTransB$.
As an example, consider the translation functions, depicted in \autoref{fig:pt_pn_translation}.
Note that, each transformation function in \autoref{fig:pt_pn_translation}, is sound by construction.
Hence, we deduce that their recursive composition is also a sound~\cite[Theorem~3.3]{DBLP:conf/bpm/Aalst00}.

\section{Translating Workflow Nets to Process Trees}
\label{sec:method}
In this section, we describe our main approach.
In~\autoref{sec:approach_overview}, we sketch the main idea of the approach.
In~\autoref{sec:ptree_nets}, we present PTree-nets, i.e., Petri nets with $rng(\labelFunc){=}\univProcessTree$, which we exploit in our approach.
In~\autoref{sec:patterns}, we present Petri net fragments, used to identify process tree operators within the net, together with a generic reduction function.
Finally, in \autoref{sec:algorithm}, we provide an algorithmic description that allows us to find process trees, including correctness proofs.

\subsection{Overview}
\label{sec:approach_overview}
The core idea of the approach, concerns searching for fragments in the given WF-net, that represent behavior that is expressible as a process tree.
The patterns we look for, bear great similarity with the translation patterns defined in \autoref{fig:pt_pn_translation}, i.e., they can be considered as a generalized \emph{reverse} of those patterns.
When we find a pattern, we replace it with a smaller net fragment that represents the process tree that was just found.
We continue to search for patterns in the reduced net, until we are not able to find any more patterns.
As we prove in \autoref{sec:algorithm}, in case the final WF-net contains just one transition, in fact, its label carries a process tree with the same labelled-language as the original WF-net.

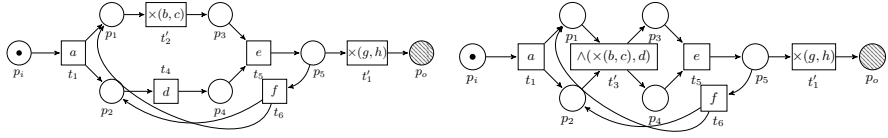
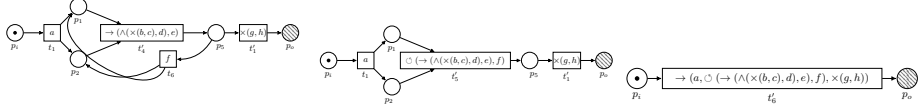
\begin{figure}[tb]
    \centering
    \begin{subfigure}[t]{0.475\textwidth}
        \centering
        \resizebox{\textwidth}{!}{
    		\begin{tikzpicture}
    		[node distance=1.35cm,
    		on grid,>=stealth',
    		bend angle=30,
    		auto,
    		every place/.style= {minimum size=6mm},
    		every transition/.style = {minimum size = 6mm}
    		]
    		
    		\node [place, tokens = 1] (start) [label=below:$\place_i$]{};
    		
    		\node [transition] (t1) [label=below:$\transition_1$, label=center:$a$, right = of start] {}
    		edge [pre] node[auto] {} (start);
    		
    		\node [place] (p1) [label=below:$\place_1$, above right = of t1] {}
    		edge [pre] node[auto] {} (t1);
    		
    		\node [transition] (t3) [label=below:$\transition'_2$, label=center: {$\times(b,c)$}, minimum width=10mm, right = of p1] {}
    		edge [pre] node[auto] {} (p1);
    		
    		\node [place] (p2) [label=below:$\place_2\ \ $, below right = of t1] {}
    		edge [pre] node[auto] {} (t1);	
    		
    		\node [transition] (t4) [label=above:$\transition_4$, label=center:$d$, right = of p2] {}
    		edge [pre] node[auto] {} (p2);
    		
    		\node [place] (p3) [label=below:$\place_3$, right = of t3] {}
    		edge [pre] node[auto] {} (t3);
    		
    		\node [place] (p4) [label=below:$\place_4$, right = of t4] {}
    		edge [pre] node[auto] {} (t4);
    		
    		\node [transition] (t5) [label=below:$\transition_5$, label=center:$e$, above right = of p4] {}
    		edge [pre] node[auto] {} (p3)
    		edge [pre] node[auto] {} (p4);
    		
    		\node [place] (p5) [label=below:$\ \ \ \place_5$, right = of t5] {}
    		edge [pre] node[auto] {} (t5);
    		
    		\node [transition] (t6) [label=below:$\ \ \transition_6$, label=center:$f$, right = of p4] {}
    		edge [pre, bend right] node[auto] {} (p5)
    		edge [post, bend left = 35] node[auto] {} (p2)
    		edge [post, bend left = 100] node[auto] {} (p1);
    		
    		\node [transition] (t7) [label=below:$\transition'_1$, label=center: {$\times(g,h)$}, minimum width=10mm, right = of p5] {}
    		edge [pre] node[auto] {} (p5);
    		
    		\node [place] (end) [label=below:$\place_o$, right = of t7, pattern=custom north west lines,hatchspread=1.5pt,hatchthickness=0.3pt,hatchcolor=gray,] {}
    		edge [pre] node[auto] {} (t7);		
    		\end{tikzpicture}
    	}
    	\caption{Result of the first two rounds of the algorithm. The first two  patterns that can be found are choice constructs, between $b$ and $c$, and, $g$ and $h$, respectively.}
    	\label{fig:application_1}
    \end{subfigure}~
    \begin{subfigure}[t]{0.475\textwidth}
        \centering
        \resizebox{\textwidth}{!}{
    		\begin{tikzpicture}
    		[node distance=1.35cm,
    		on grid,>=stealth',
    		bend angle=30,
    		auto,
    		every place/.style= {minimum size=6mm},
    		every transition/.style = {minimum size = 6mm}
    		]
    		
    		\node [place, tokens = 1] (start) [label=below:$\place_i$]{};
    		
    		\node [transition] (t1) [label=below:$\transition_1$, label=center:$a$, right = of start] {}
    		edge [pre] node[auto] {} (start);
    		
    		\node [place] (p1) [label=below:$\place_1$, above right = of t1] {}
    		edge [pre] node[auto] {} (t1);

    		\node [place] (p2) [label=below:$\place_2\ \ $, below right = of t1] {}
    		edge [pre] node[auto] {} (t1);	
    		
    		\node [transition] (t4) [label=below:$\transition'_3$,  label=center: {$\wedge(\times(b,c),d)$}, minimum width=20mm, above right = of p2] {}
    		edge [pre] node[auto] {} (p2)
    		edge [pre] node[auto] {} (p1);
    		
    		\node [place] (p3) [label=below:$\place_3$, above right = of t4] {}
    		edge [pre] node[auto] {} (t4);
    		
    		\node [place] (p4) [label=below:$\place_4$, below right = of t4] {}
    		edge [pre] node[auto] {} (t4);
    		
    		\node [transition] (t5) [label=below:$\transition_5$, label=center:$e$, above right = of p4] {}
    		edge [pre] node[auto] {} (p3)
    		edge [pre] node[auto] {} (p4);
    		
    		\node [place] (p5) [label=below:$\ \ \ \place_5$, right = of t5] {}
    		edge [pre] node[auto] {} (t5);
    		
    		\node [transition] (t6) [label=below:$\ \ \transition_6$, label=center:$f$, right = of p4] {}
    		edge [pre, bend right] node[auto] {} (p5)
    		edge [post, bend left = 35] node[auto] {} (p2)
    		edge [post, bend left = 100] node[auto] {} (p1);
    		
    		\node [transition] (t7) [label=below:$\transition'_1$, label=center: {$\times(g,h)$}, minimum width=10mm, right = of p5] {}
    		edge [pre] node[auto] {} (p5);
    		
    		\node [place] (end) [label=below:$\place_o$, right = of t7, pattern=custom north west lines,hatchspread=1.5pt,hatchthickness=0.3pt,hatchcolor=gray,] {}
    		edge [pre] node[auto] {} (t7);		
    		\end{tikzpicture}
    	}
    	\caption{Result of the third round of the algorithm on the running example. We find a parallel construct between transition $\transition'_2$ and $\transition_4$.}
    	\label{fig:application_2}
    \end{subfigure}
    \begin{subfigure}[t]{0.325\textwidth}
        \centering
        \resizebox{\textwidth}{!}{
    		\begin{tikzpicture}
    		[node distance=1.35cm,
    		on grid,>=stealth',
    		bend angle=30,
    		auto,
    		every place/.style= {minimum size=6mm},
    		every transition/.style = {minimum size = 6mm}
    		]
    		
    		\node [place, tokens = 1] (start) [label=below:$\place_i$]{};
    		
    		\node [transition] (t1) [label=below:$\transition_1$, label=center:$a$, right = of start] {}
    		edge [pre] node[auto] {} (start);
    		
    		\node [place] (p1) [label=below:$\place_1$, above right = of t1] {}
    		edge [pre] node[auto] {} (t1);

    		\node [place] (p2) [label=below:$\place_2\ \ $, below right = of t1] {}
    		edge [pre] node[auto] {} (t1);
    		
    		\node [] (rbp1) [below right = of p1] {};
    		
    		\node [transition] (t4) [label=below:$\transition'_4$,  label=center: {$\to(\wedge(\times(b,c),d),e)$}, minimum width=30mm, right = of rbp1] {}
    		edge [pre] node[auto] {} (p2)
    		edge [pre] node[auto] {} (p1);
    		
    		\node [] (rt4) [right = of t4] {};
    		
    		\node [place] (p5) [label=below:$\ \ \ \place_5$, right = of rt4] {}
    		edge [pre] node[auto] {} (t4);
    		
    		\node [transition] (t6) [label=below:$\ \ \transition_6$, label=center:$f$, right = of p4] {}
    		edge [pre, bend right] node[auto] {} (p5)
    		edge [post, bend left = 35] node[auto] {} (p2)
    		edge [post, bend left = 100] node[auto] {} (p1);
    		
    		\node [transition] (t7) [label=below:$\transition'_1$, label=center: {$\times(g,h)$}, minimum width=10mm, right = of p5] {}
    		edge [pre] node[auto] {} (p5);
    		
    		\node [place] (end) [label=below:$\place_o$, right = of t7, pattern=custom north west lines,hatchspread=1.5pt,hatchthickness=0.3pt,hatchcolor=gray,] {}
    		edge [pre] node[auto] {} (t7);		
    		\end{tikzpicture}
    	}
    	\caption{Result of the fourth round of the algorithm. We find a sequential construct.}
    	\label{fig:application_3}
    \end{subfigure}~
    \begin{subfigure}[t]{0.325\textwidth}
        \centering
        \resizebox{\textwidth}{!}{
    		\begin{tikzpicture}
    		[node distance=1.35cm,
    		on grid,>=stealth',
    		bend angle=30,
    		auto,
    		every place/.style= {minimum size=6mm},
    		every transition/.style = {minimum size = 6mm}
    		]
    		
    		\node [place, tokens = 1] (start) [label=below:$\place_i$]{};
    		
    		\node [transition] (t1) [label=below:$\transition_1$, label=center:$a$, right = of start] {}
    		edge [pre] node[auto] {} (start);
    		
    		\node [place] (p1) [label=below:$\place_1$, above right = of t1] {}
    		edge [pre] node[auto] {} (t1);

    		\node [place] (p2) [label=below:$\place_2\ \ $, below right = of t1] {}
    		edge [pre] node[auto] {} (t1);
    		
    		\node [] (rbp1) [below right = of p1] {};
    		
    		\node [transition] (t4) [label=below:$\transition'_5$,  label=center: {$\circlearrowleft(\to(\wedge(\times(b,c),d),e),f)$}, minimum width=40mm, right = of rbp1] {}
    		edge [pre] node[auto] {} (p2)
    		edge [pre] node[auto] {} (p1);
    		
    		\node [] (rt4) [right = of t4] {};
    		
    		\node [place] (p5) [label=below:$\ \ \ \place_5$, right = of rt4] {}
    		edge [pre] node[auto] {} (t4);
    		
    		\node [transition] (t7) [label=below:$\transition'_1$, label=center: {$\times(g,h)$}, minimum width=10mm, right = of p5] {}
    		edge [pre] node[auto] {} (p5);
    		
    		\node [place] (end) [label=below:$\place_o$, right = of t7, pattern=custom north west lines,hatchspread=1.5pt,hatchthickness=0.3pt,hatchcolor=gray,] {}
    		edge [pre] node[auto] {} (t7);		
    		\end{tikzpicture}
    	}
    	\caption{Result of the fifth round of the algorithm. We find a loop construct.}
    	\label{fig:application_4}
    \end{subfigure}
    %
    %
    %
    %
    %
    %
    %
    \begin{subfigure}[t]{0.325\textwidth}
        \centering
        \resizebox{\textwidth}{!}{
    		\begin{tikzpicture}
    		[node distance=2cm,
    		on grid,>=stealth',
    		bend angle=30,
    		auto,
    		every place/.style= {minimum size=6mm},
    		every transition/.style = {minimum size = 6mm}
    		]
    		
    		\node [place, tokens = 1] (start) [label=below:$\place_i$]{};
    		
    		\node [] (rstart) [right = of start] {};
    		
    		\node [transition] (t1) [label=below:$\transition'_6$, label=center: {$\to(a,\circlearrowleft(\to(\wedge(\times(b,c),d),e),f),\times(g,h))$}, minimum width=65mm, right = of rstart] {}
    		edge [pre] node[auto] {} (start);
    		
    		\node [] (rt1) [right = of t1] {};
    		
    		\node [place] (end) [label=below:$\place_o$, right = of rt1, pattern=custom north west lines,hatchspread=1.5pt,hatchthickness=0.3pt,hatchcolor=gray,] {}
    		edge [pre] node[auto] {} (t1);		
    		\end{tikzpicture}
    	}
    	\caption{Result of the final round of the algorithm. We find a sequence construct.}
    	\label{fig:application_6}
    \end{subfigure}
    
    \caption{Application of the algorithm on the running example, i.e., $\wfNet_1$. The label of $\transition'_6$, i.e., $\labelFuncPTWF(\transition'_6)$, depicted in \autoref{fig:application_6}, is the resulting process tree. The resulting process tree, i.e., $\to(a,\circlearrowleft(\to(\wedge(\times(b,c),d),e),f),\times(g,h))$, is equal to \autoref{fig:example_tree}.}
    \label{fig:application}
\end{figure}

Consider \autoref{fig:application}, in which we sketch the basic idea of the algorithm, applied on the example WF-net $\wfNet_1$ (\autoref{fig:example_petri_net}).
First, the algorithm detects two \emph{choice constructs}, i.e., one between the transitions labelled $b$ and $c$, and, one between the transitions labelled $g$ and $h$.
The algorithm replaces the fragments by means of two new transitions, carrying labels $\times(b,c)$ and $\times(g,h)$ respectively (\autoref{fig:application_1}).
Subsequently, a parallel construct is detected, i.e., between the transitions labelled $\times(b,c)$ and $d$.
Again, the pattern is replaced (\autoref{fig:application_2}).
A sequential pattern is detected and replaced (\autoref{fig:application_3}), after which a loop construct is detected (\autoref{fig:application_4}).
The resulting process tree, i.e., carried by the remaining transition in \autoref{fig:application_6}, $\to(a,\circlearrowleft(\to(\wedge(\times(b,c),d),e),f),\times(g,h))$, is equal to \autoref{fig:example_tree}.

\subsection{PTree-Nets and their Unfolding}
\label{sec:ptree_nets}
The core idea of the proposed algorithm presented, is finding Petri net fragments in the WF-net, that represent behavior equivalent to a process tree.
As illustrated in \autoref{sec:approach_overview}, the patterns found in the WF-net are replaced by transitions with a label carrying a corresponding process tree.
In the upcoming section, we present four different fragment characterizations, corresponding to the basic process tree operators considered. 
However, in this section, we first briefly present PTree-nets, i.e., a trivial extension of Petri nets, in which labels are process trees.
\begin{definition}[Process Tree-Labelled Petri-net (PTree-net)]
\label{def:ptwf_net}
Let $\univProcessTree$ denote the universe of process trees.
Let $\tilde{\places}$ denote a set of places, let $\tilde{\transitions}$ denote a set of transitions and let $\tilde{\netArcs}{\subseteq}(\tilde{\places}{\times}\tilde{\transitions}){\cup}(\tilde{\transitions}{\times}\tilde{\places})$ denote the arc relation.
Let $\labelFuncPTWF{\colon}\tilde{\transitions}{\to}\univProcessTree$.
Tuple $\ptNet{=}(\tilde{\places},\tilde{\transitions},\tilde{\netArcs},\labelFuncPTWF)$ is a Process Tree-Labelled Petri net (PTree-net).
$\univPtNet$ denotes the universe of PTree-nets.
\end{definition}
Given $\ptNet{\in}\univPtNet$, we have $\labelFuncPTWF\left(\langNet\left(\ptNet,\marking,\marking'\right)\right){\in}\univProcessTree^*$, and,  $\langPT\left(\labelFuncPTWF\left(\langNet\left(\ptNet,\marking,\marking'\right)\right)\right){\in}\labels^*$, i.e., the definition of $\langPT$ ignores $\silent{\notin}{\labels}$.\footnote{
Observe: $\langNetVis\left(\ptNet,\marking,\marking'\right){=}\labelFuncPTWF\left(\langNet\left(\ptNet,\marking,\marking'\right)\right)_{\downarrow_{\labels}}{=}\langPT\left(\labelFuncPTWF\left(\langNet\left(\ptNet,\marking,\marking'\right)\right)\right)$.}
Clearly, since PTree-nets extend the labelling function to $\univProcessTree$, PTree-System-nets, and, PTree-WF-nets are readily defined.
We let $\univPtSystemNet$ and $\univPtWfNet$ represent their respective universes.

Since a PTree-net contains process trees as labels, which can be translated into a Petri net fragment, we define a PTree-net \emph{unfolding}, cf. \autoref{def:ptwf_unfolding}, which maps a PTree-net onto a corresponding conventional Petri net.
\begin{definition}[PTree-net Unfolding]
\label{def:ptwf_unfolding}
A PTree-net unfolding $\ptWfNetUnfolding{\colon}\univPtNet{\to}\univNets$ is a function where, given $\ptNet{=}(\tilde{\places},\tilde{\transitions},\tilde{\netArcs},\labelFuncPTWF){\in}\univPtNet$, ${\ptWfNetUnfolding(\ptNet){=}(\places,\transitions,\netArcs,\labelFunc)}$, with: 
\begin{enumerate}
    \item[] Let $\ptToWfNetTranslator(\labelFuncPTWF(\tilde{\transition})){=}({\places}_{\tilde{\transition}},{\transitions}_{\tilde{\transition}},\netArcs_{\tilde{\transition}},\place_{i_{\tilde{\transition}}},\place_{o_{\tilde{\transition}}},\labelFunc_{\tilde{\transition}}), \ptToWfNetTranslatorTransB(\labelFuncPTWF(\tilde{\transition})){=}(\hat{\places}_{\tilde{\transition}},\hat{\transitions}_{\tilde{\transition}},\hat{\netArcs}_{\tilde{\transition}},\hat{\labelFunc}_{\tilde{\transition}}), \forall{\tilde{\transition}{\in}\tilde{\transitions}}$,
    \item $\places{=}\tilde{\places}{\cup}\bigcup\limits_{\tilde{\transition}{\in}\tilde{\transitions}}{\hat{\places}_{\tilde{\transition}}}$,
    \item $\transitions{=}\bigcup\limits_{\tilde{\transition}{\in}\tilde{\transitions}}{\hat{\transitions}_{\tilde{\transition}}}$,
    \item $\netArcs{=}\bigcup\limits_{\tilde{\transition}{\in}\tilde{\transitions}}\hat{\netArcs}_{\tilde{\transition}}{\cup}\bigcup\limits_{\tilde{\transition}{\in}\tilde{\transitions}}\{(\place,\tilde{\transition})\mid \place{\in}{\bullet}{\tilde{\transition}}{\wedge}\tilde{\transition}{\in}\place_{i_{\tilde{\transition}}}{\bullet}\}{\cup}\bigcup\limits_{\tilde{\transition}{\in}\tilde{\transitions}}\{(\tilde{\transition},\place)\mid \place{\in}{\tilde{\transition}}{\bullet}{\wedge}\tilde{\transition}{\in}{\bullet}{\place_{o_{\tilde{\transition}}}}\}$,
    \item $\labelFunc{=}\bigcup\limits_{\tilde{\transition}{\in}\tilde{\transitions}}\hat{\labelFunc}_{\tilde{\transition}}$.\footnote{Since functions are binary Cartesian products, we write set operations here.}
\end{enumerate}
\end{definition}
Note that, the WF-net in \autoref{fig:example_petri_net}, is the unfolding of all PTree-WF-nets in \autoref{fig:application}.

\subsection{Pattern Reduction}
\label{sec:patterns}
In this section, we describe four patterns, used to identify and replace process tree behavior.
Furthermore, we propose a corresponding overarching reduction function, which shows how to reduce a PTree-WF-net containing any of these patterns.
However, first, we present the general notion of a \emph{feasible pattern}.
\begin{definition}[Feasible $\processTreeOp$-Pattern]
\label{def:feasible_pattern}
Let $\ptNet{=}(\tilde{\places},\tilde{\transitions},\tilde{\netArcs},\labelFuncPTWF){\in}\univPtNet$, let\\ $\splitatcommas{\net{=}(\places,\transitions{=}\left\{\transition_1,...,\transition_n\right\},\netArcs,\labelFuncPTWF|_{\transitions}){\sqsubseteq}\ptNet}$ and let $\marking_i,\marking_f{\in}\univMultisets(\places)$ ($n{\geq}1$).
Given $\processTreeOp{\in}\univProcessTreeOperators$, $\systemNet{=}(\net,\marking_i,\marking_f){\in}\univPtSystemNet$ is a feasible $\processTreeOp$-pattern, written  $\pnptPattern_{\processTreeOp}(\ptNet,\systemNet)$, iff:
\begin{enumerate}
    \item $\systemNet$ is \emph{weakly connected},
    \item $\marking_i(\place){\leq}1{\wedge}\marking_f(\place){\leq}1$, $\forall p{\in}\places$,
    \item $\langPT\left(\processTreeOp\left(\labelFuncPTWF\left(\transition_1\right),...,\labelFuncPTWF\left(\transition_n\right)\right)\right){=}\langNetVis\left(\ptWfNetUnfolding\left(\net\right),\marking_i,\marking_f\right)$. 
\end{enumerate}
\end{definition}
In the upcoming paragraphs, we characterize an instantiation of a feasible $\processTreeOp$-pattern, for each operator considered.

\subsubsection{Sequential Pattern}
The $\to$-operator, i.e., ${\to}\left(\processTree_1,...,\processTree_n\right)$, describes sequential behavior, hence, any subnet describing \emph{strictly sequential behavior}, describes the same language.
If a transition $\transition_1$ always, uniquely, enables transition $\transition_2$, which in turn enables transition $\transition_3,...,\transition_n$, and, whenever $\transition_1$ has fired, the only way to consume all tokens from $\transition_1{\bullet}$, is by means of firing $\transition_2$, and similarly, the only way to consume all tokens from $\transition_2{\bullet}$, is by means of firing $\transition_3$, etc., then $\transition_1,...,\transition_n$ are in a sequential relation.
We visualize the $\to$-pattern in the left-hand side of \autoref{fig:seq_pattern}.
\begin{definition}[$\to$-Pattern]
\label{def:seq_pattern}
Let $\ptNet{=}(\tilde{\places},\tilde{\transitions},\tilde{\netArcs},\labelFuncPTWF){\in}\univPtNet$ and let $\left\{\transition_1,...,\transition_n\right\}{\subseteq}\tilde{\transitions}$ ($n{\geq}2$). 
If and only if:
\begin{enumerate}
    \item $\forall{1{\leq}i{<}n}\left(|\transition_i{\bullet}|{\geq}1{\wedge}\transition_i{\bullet}{=}{\bullet}{\transition_{i+1}}\right)$; transition $\transition_i$ enables $\transition_{i+1}$,
    \item $\forall{1{\leq}i{<}n}\left(\forall{\place{\in}\transition_i{\bullet}}\left({\bullet}{\place}{=}\{\transition_i\}{\wedge}{\place}{\bullet}{=}\{\transition_{i+1}\}\right)\right)$; enabling is unique,
    \item $\bigcap\limits_{\transition{\in}\transitions}{\transition{\bullet}}{=}\emptyset$; self-loops are not allowed,
\end{enumerate}
then, system net $SN{=}(\net{=}(\places,\transitions,\netArcs,\labelFuncPTWF|_{\transitions}),{\bullet}\transition_1,\transition_n{\bullet})$, with $\places{=}{\bullet}\transition_1{\cup}{\bullet}\transition_2{\cup}{\cdots}{\bullet}\transition_n{\cup}\transition_n{\bullet}$, $\transitions{=}\{\transition_1,...,\transition_n\}$, $\netArcs{=}\{(x,y){\in}\tilde{\netArcs}\mid y{\in}{\transitions}{\vee}x{=}\transition_n\}$, is a feasible $\to$-pattern.
\end{definition}

\subsubsection{Exclusive Choice Pattern}
\begin{figure}[tbp]
    \centering
    \begin{subfigure}[t]{\textwidth}
        \centering
        \includegraphics[width=1\textwidth]{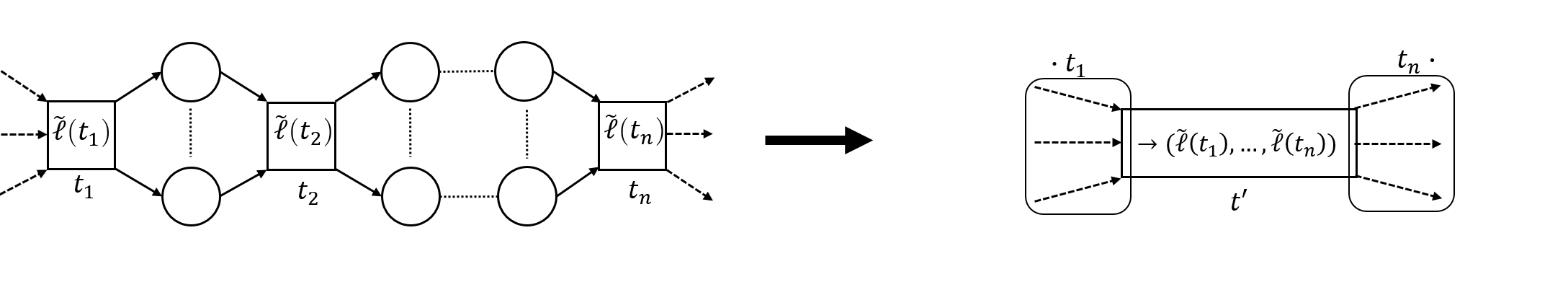}
        \caption{Visualization of the $\to$-pattern reduction. The post-set of each transition $\transition_i$ acts as the pre-set of $\transition_{i+1}$ ($1{\leq}i{<}n$). The replacing transition inherits ${\bullet}\transition_1$ and $\transition_n{\bullet}$.}
        \label{fig:seq_pattern} 
    \end{subfigure}
    \begin{subfigure}[t]{\textwidth}
        \centering
        \includegraphics[width=1\textwidth]{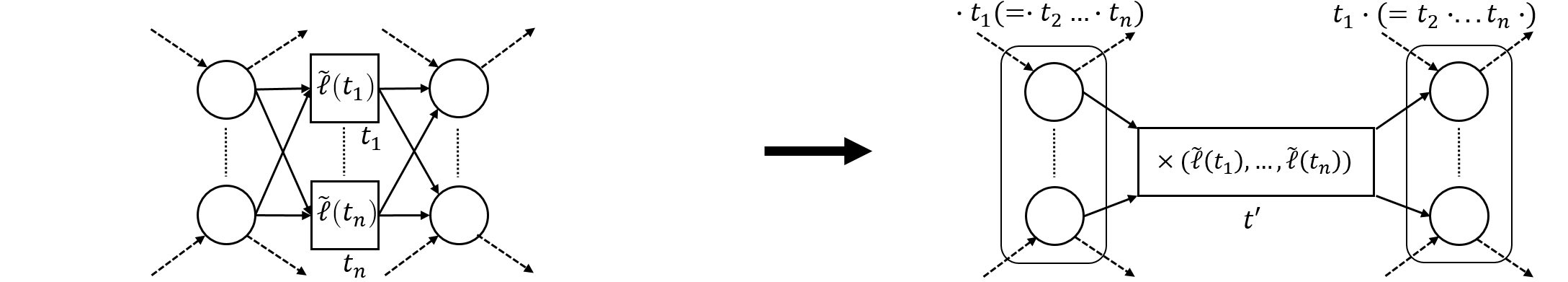}
        \caption{Visualization of the $\times$-pattern reduction. All transitions in the pattern share the same pre- and post-set. The replacing transition inherits said pre- and post-set.}
        \label{fig:xor_pattern} 
    \end{subfigure}
    \begin{subfigure}[t]{\textwidth}
        \centering
        \includegraphics[width=1\textwidth]{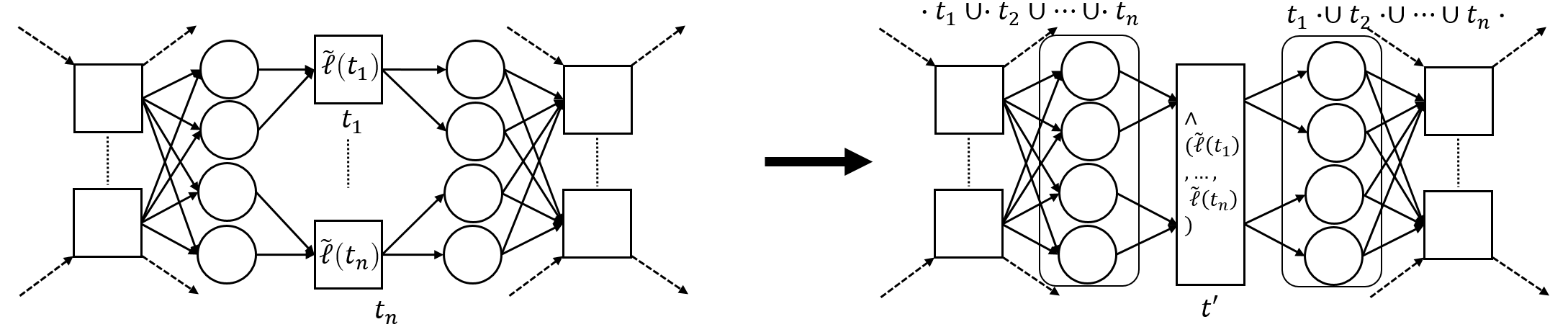}
        \caption{Visualization of the $\wedge$-pattern reduction. Transitions $\transition_1,...,\transition_n$ have disjunct pre-sets, yet, their pre-sets have the exact same pre-sets. The same holds for the post-sets of transitions $\transition_1,..,\transition_n$. The replacing transition inherits all pre- and post-sets.}
        \label{fig:par_pattern} 
    \end{subfigure}
    \begin{subfigure}[t]{\textwidth}
        \centering
        \includegraphics[width=1\textwidth]{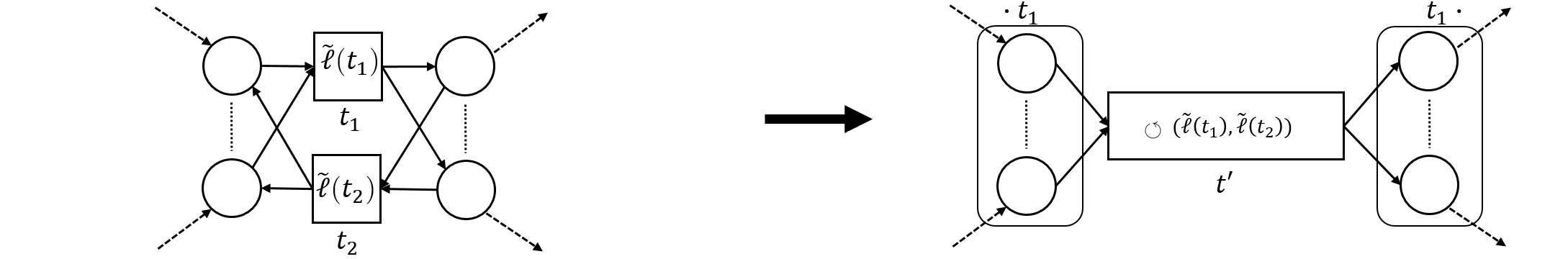}
        \caption{Visualization of the $\circlearrowleft$-pattern reduction. The pre-set of transition $\transition_1$ equals the post-set of $\transition_2$ and vice-versa. The replacing transition inherits the pre- and post-set of transition $\transition_1$.}
        \label{fig:loop_pattern} 
    \end{subfigure}
    \caption{Visualization of the reductions considered in this paper. Solid arcs are required to be present, dashed arcs are allowed to be present, i.e., absence of dashed arcs, yet presence of solid arcs, implies that only the solid arcs are allowed.}
    \label{fig:reductions}
\end{figure}

The $\times$-operator, i.e., $\times(\processTree_1, ..., \processTree_n)$, describes ``execute either one of $\processTree_1, ..., \processTree_n$''.
In terms of a Petri net fragment, transitions $\transition_1,...,\transition_n$ are in an exclusive choice pattern if their pre and post-sets are equal.
Consider \autoref{fig:xor_pattern}, in which we schematically depict the $\times$-pattern.
\begin{definition}[$\times$-Pattern]
\label{def:xor_pattern}
Let $\ptNet{=}(\tilde{\places},\tilde{\transitions},\tilde{\netArcs},\labelFuncPTWF){\in}\univPtNet$ and let $\left\{\transition_1,...,\transition_n\right\}{\subseteq}\tilde{\transitions}$ ($n{\geq}2$). 
If and only if:
\begin{enumerate}
    \item ${\bullet}\transition_1{=}{\bullet}\transition_2{=}{\cdots}{=}{\bullet}\transition_n$; all pre-sets are shared among the members of the pattern,
    \item $\transition_1{\bullet}{=}\transition_2{\bullet}{=}{\cdots}{=}\transition_n{\bullet}$; all post-sets are shared among the members of the pattern,
    \item $\forall{1{\leq}i{\leq}n}\left({\bullet}\transition_i{\neq}\transition_i{\bullet}\right)$; self-loops are not allowed,
\end{enumerate}
then, system net $SN{=}\left(\net{=}\left(\places,\transitions,\netArcs,\labelFuncPTWF|_{\transitions}\right),{\bullet}\transition_1,\transition_1{\bullet}\right)$, with $\places{=}{\bullet}\transition_1{\cup}\transition_1{\bullet}$, $\transitions{=}\{\transition_1,...,\transition_n\}$, $\netArcs{=}\{(x,y){\in}\tilde{\netArcs}\mid x{\in}\transitions{\vee}y{\in}\transitions\}$, is a feasible $\times$-pattern.
\end{definition}

\subsubsection{Parallel Pattern}
The parallel pattern is the most complicated pattern.
In such a fragment, inference between its transitions is possible.
This is achieved by requiring that the pre-sets and post-sets of the transitions do not have any overlap.
Furthermore, the pre-set of the transition's pre-set places needs to be shared by all of these places, and, symmetrically, the post-set of the transition's post-set places needs to be shared by all of these places.
That is, the enabling of the transitions in the pattern needs to be the same, and, their post-set should jointly block any further action, until all places in their joint post-set are marked.
\begin{definition}[$\wedge$-Pattern]
\label{def:par_pattern}
Let $\ptNet{=}(\tilde{\places},\tilde{\transitions},\tilde{\netArcs},\labelFuncPTWF){\in}\univPtNet$ and let $\transitions{=}\left\{\transition_1,...,\transition_n\right\}{\subseteq}\tilde{\transitions}$ ($n{\geq}2$).
If and only if:
\begin{enumerate}
    \item $\forall{1{\leq}i{<}j{\leq}n}\left({\bullet}{\transition_i}{\cap}{\bullet}{\transition_j}{=}\emptyset\right)$; no interaction between the member's pre-sets,
    \item $\forall{1{\leq}i{<}j{\leq}n}\left({\transition_i}{\bullet}{\cap}{\transition_j}{\bullet}{=}\emptyset\right)$; no interaction between the member's post-sets,
    \item $\forall{1{\leq}i{\leq}n}\left(\forall{\place{\in}{\bullet}{\transition_i}}\left({\place}{\bullet}{=}\left\{\transition_i\right\}\right)\right)$; pre-set places uniquely connect to a member,
    \item $\forall{1{\leq}i{\leq}n}\left(\forall{\place{\in}{\transition_i}{\bullet}}\left({\bullet}{\place}{=}\left\{\transition_i\right\}\right)\right)$; post-set places uniquely connect to a member,
    \item $\forall{\place}{\in}{\bullet}\transitions\left({\bullet}{\place}{\cap}\{\transition_1,...,\transition_n\}{=}\emptyset\right)$; members do not influence other members,
    \item $\forall{\place,\place'}{\in}{\bullet}\transitions\left({\bullet}{\place}{=}{\bullet}{\place'}\right)$; member's pre-sets share their pre-set,
    \item $\forall{\place}{\in}\transitions{\bullet}\left({\place}{\bullet}{\cap}\{\transition_1,...,\transition_n\}{=}\emptyset\right)$; member firing does not affect other members,
    \item $\forall{\place,\place'}{\in}\transitions{\bullet}\left({\place}{\bullet}{=}{\place'}{\bullet}\right)$; member's post-sets share their post-set,
    \item $\forall{\transition,\transition'}{\in}\bullet\left(\bullet\transitions\right)\left({\bullet}\transition{=}{\bullet}\transition'\right)$; pre-sets of enablers are equal,
    \item $\forall{\transition,\transition'}{\in}\left(\transitions{\bullet}\right){\bullet}\left(\transition{\bullet}{=}\transition'{\bullet}\right)$; post-sets of enablers are equal,
\end{enumerate}
then, system net $SN{=}\left(\net{=}\left(\places,\transitions,\netArcs,\labelFuncPTWF|_{\transitions}\right),{\bullet}\left({\bullet}\left({\bullet}\transitions\right)\right),\left(\left(\transitions{\bullet}\right){\bullet}\right){\bullet}\right)$, with\\ $\places{=}{\bullet}\transitions{\cup}{\bullet}\left({\bullet}\left({\bullet}\transitions\right)\right){\cup}\transitions{\bullet}{\cup}\left(\left(\transitions{\bullet}\right){\bullet}\right){\bullet}$, $\transitions{=}\transitions$, $\splitatcommas{\netArcs{=}\{(x,y){\in}\tilde{\netArcs} \mid  x{\in}{\bullet}\left({\bullet}\left({\bullet}\transitions\right)\right) \cup {\bullet}\transitions{\cup}\transitions{\bullet}{\vee}y{\in}\left(\left(\transitions{\bullet}\right){\bullet}\right){\bullet}{\cup}{\bullet}\transitions{\cup}\transitions{\bullet}\}}$ is a feasible $\wedge$-pattern.
\end{definition}

\subsubsection{Loop Pattern}
The final operator we consider, is the $\circlearrowleft$-operator, i.e., the only operator with just two children.
Hence, the fragments representing a loop pattern in the Ptree-net consist of two transitions.
Consider the left-hand side of \autoref{fig:loop_pattern}, in which we schematically depict the loop pattern fragment.

\begin{definition}[$\circlearrowleft$-Pattern]
\label{def:loop_pattern}
Let $\ptNet{=}(\tilde{\places},\tilde{\transitions},\tilde{\netArcs},\labelFuncPTWF){\in}\univPtNet$ and let $\transition_1{\neq}\transition_2{\in}\tilde{\transitions}$. Iff:
\begin{enumerate}
    \item ${\bullet}\transition_1{=}\transition_2{\bullet}$; pre-set of $\transition_1$ is the post-set of $\transition_2$,
    \item $\transition_1{\bullet}{=}{\bullet}\transition_2$; pre-set of $\transition_2$ is the post-set of $\transition_1$,
    \item ${\bullet}\transition_1{\cap}{\bullet}\transition_2{=}\emptyset$; pre-sets are disjunct,
    \item $\transition_1{\bullet}{\cap}\transition_2{\bullet}{=}\emptyset$; post-sets are disjunct.
\end{enumerate}
then, system net $SN{=}\left(\net{=}\left(\places,\transitions,\netArcs,\labelFuncPTWF|_{\{\transition_1,\transition_2\}}\right),{\bullet}\transition_1,\transition_1{\bullet}\right)$, with $\places{=}{\bullet}\transition_1{\cup}\transition_1{\bullet}$, $\transitions{=}\{\transition_1,\transition_2\}$, $\netArcs{=}\{(x,y){\in}\tilde{\netArcs}\mid x{\in}\{\transition_1,\transition_2\}{\vee}y{\in}\{\transition_1,\transition_2\}\}$ is a feasible $\circlearrowleft$-pattern.
\end{definition}

Observe that, for each of the patterns presented here, it is easy to see that the requirements posed in \autoref{def:feasible_pattern}, hold.

\subsubsection{Pattern Reduction}
The reduction rules for the patterns, i.e., defined in the previous paragraphs, as depicted in the right-hand side of \autoref{fig:reductions}, are very similar.
Except for the $\to$-pattern, we ``copy'' all places in the new Ptree-net, i.e., for the $\to$-pattern, we remove the places inter-connecting transitions $\transition_1,...,\transition_n$.
The transitions involved in the pattern (and connecting arcs) are removed.
A new transition is inserted, with a label depending on the pattern found.
The connecting arcs of the newly added transition, differ slightly per pattern.

\begin{definition}[Pattern Reduction]
\label{def:reduction}
Let $\processTreeOp{\in}\{\to,\times,\wedge,\circlearrowleft\}$, let $\splitatcommas{\ptNet{=}(\tilde{\places},\tilde{\transitions},\tilde{\netArcs},\labelFuncPTWF){\in}\univPtNet}$, let $\net{=}(\places,\transitions{=}\{\transition_1,...,\transition_n\},\netArcs,\labelFuncPTWF|_{\transitions}){\sqsubseteq}\ptNet$, let $\marking_i,\marking_f{\subseteq}\univMultisets(\places)$ and let $\systemNet{=}\left(\net,\marking_i,\marking_f\right)$ s.t. $\pnptPattern_{\processTreeOp}(\ptNet,\systemNet)$.
We let ${\pnptRedution_{\processTreeOp}(\ptNet,\systemNet){=}\ptNet'{=}(\tilde{\places}',\tilde{\transitions}',\tilde{\netArcs}',\labelFuncPTWF')}$ denote the $\pnptRedution_{\processTreeOp}(\ptNet,\systemNet)$-reduced PTree-net, with, for $\tilde{\transition}'{\notin}\tilde{\transitions}$:
\begin{itemize}
\item $\tilde{\places}'{=}\begin{cases}\tilde{\places}{\setminus}\bigcup\limits_{i{=}1}^{n-1}\transition_i{\bullet} & if\ \processTreeOp{=}\to\\ \tilde{\places} & otherwise \end{cases}$,
\item $\tilde{\transitions}'{=}\left(\tilde{\transitions}{\setminus}\transitions\right){\cup}\{\tilde{\transition}'\}$,
\item $\tilde{\netArcs}'{=}\left(\tilde{\netArcs}{\setminus}\bigcup\limits_{i{=}1}^{n}\left(\{(\transition_i,\place) \mid \place{\in}\transition_i{\bullet}\}{\cup}\{(\place,\transition_i) \mid \place{\in}{\bullet}\transition_i\}\right)\right){\cup}$
\[
\begin{cases}\{(\place,\tilde{\transition}') \mid \place{\in}\bullet{\transition_1}\}{\cup}\{(\tilde{\transition}',\place) \mid \place{\in}\transition_n{\bullet}\} & if\ \processTreeOp{=}\to\\
\bigcup\limits_{i{=}1}^{n}\left(\{(\place,\tilde{\transition}') \mid \place{\in}\bullet{\transition_i}\}{\cup}\{(\tilde{\transition}',\place) \mid \place{\in}\transition_i{\bullet}\}\right) & if \processTreeOp{\in}\{\times,\wedge\}\\
\{(\place,\tilde{\transition}') \mid \place{\in}\bullet{\transition_1}\}{\cup}\{(\tilde{\transition}',\place) \mid \place{\in}\transition_1{\bullet}\} & if\ \processTreeOp{=}\circlearrowleft\end{cases},\]
\item $\labelFuncPTWF'{=}\labelFuncPTWF|_{\tilde{\transitions}{\setminus}\transitions}{\cup}\{(\tilde{\transition}',\processTreeOp(\labelFuncPTWF(\transition_1),...,\labelFuncPTWF(\transition_n)))\}$.
\end{itemize}
\end{definition}

\subsection{Algorithm}
\label{sec:algorithm}
\begin{wrapfigure}{r}{0.55\textwidth}
\vspace{-25pt}
\begin{algorithm}[H]
\SetKwInOut{Input}{input}
\SetKwInOut{Output}{output}
\Input{$\wfNet{=}\left(\places,\transitions,\netArcs,\place_i,\place_o,\labelFunc\right){\in}\univWfNets$}
\Output{$\ptWfNet{=}(\tilde{\places},\tilde{\transitions},\tilde{\netArcs},\place_i,\place_o,\labelFuncPTWF){\in}\univPtWfNet$}
 $\ptNet{\gets}(\places, \transitions, \netArcs, \labelFunc) $\;
 \While{$\exists\ \systemNet{\in}\univPtSystemNet$ s.t. ${\pnptPattern_{\processTreeOp}(\ptNet,\systemNet)}$ for $\processTreeOp{\in}\{\to,\times,\wedge,\circlearrowleft\}$}{
  $\ptNet \gets \pnptRedution_{\processTreeOp}(\ptNet,\systemNet)$\;\label{alg:line:reduction}
 }
 \Return $(\net,\place_i,\place_o,\labelFuncPTWF)$\;
 \caption{WF-net Reduction}
 \label{alg:reduction}
\end{algorithm}
\vspace{-10pt}
\end{wrapfigure}
Here, we present an algorithm that translates a WF-net into a process tree.
We prove that, if the algorithm finds a process tree, the input WF-net is sound.
Moreover, we show that the language of the input WF-net equals the language of the process tree found.

Consider \autoref{alg:reduction}, in which we present an algorithmic description of the reduction algorithm. 
As an input, the algorithm needs any WF-net $\wfNet$. 
Initially, the elements of $\wfNet$, excluding the initial and final place, are copied into variable $\ptNet$.
In case a pattern of the form $\pnptPattern_{\processTreeOp}(\ptNet,\systemNet)$ is found in $\ptNet$, the corresponding reduction $\pnptRedution_{\processTreeOp}(\ptNet,\systemNet)$ is applied (\autoref{alg:line:reduction}).
If no more pattern is found, the algorithm returns $(\net,\place_i,\place_o,\labelFuncPTWF)$.

In case the obtained PTree-WF-net consists of just one transition, i.e., connected to place $\place_i$ (incoming) and place $\place_o$ (outgoing), cf. \autoref{fig:application_6}, the label of the transition represents a process tree, describing the same language as the original WF-net.
Furthermore, we are able to conclude that the original WF-net is, in fact, a sound WF-net.
We prove these observations in \autoref{th:algo_lang_eq}, however, prior to this, we first present two useful lemmas.
In \autoref{lemma:reduction_sound}, we prove that the proposed reduction rules are bidirectionally soundness preserving, i.e., if a PTree-WF-net is sound, then also the reduced PTree-WF-net is sound, and, vice versa.
In \autoref{lemma:reduc_lang}, we prove that, if we are able, from the initial marking $[\place_i]$, to enable the observed fragment (enabling differs per fragment), then, the language of the original net and the reduced net is equal (and vice versa).
Observe that, trivially, the reduction rules applied on a PTree-WF-net, yield a PTree-WF-net, i.e., none of the requirements of~\autoref{def:wf_net}, are violated on the resulting net.

\begin{lemma}[Pattern Reduction is Soundness Preserving]
\label{lemma:reduction_sound}
Let $\processTreeOp{\in}\{\to,\times,\wedge,\circlearrowleft\}$, let $\ptWfNet{=}(\tilde{\places},\tilde{\transitions},\tilde{\netArcs},\tilde{\place}_i,\tilde{\place}_o,\labelFuncPTWF){\in}\univPtWfNet$, let $\systemNet{\in}\univPtSystemNet$, s.t., $\pnptPattern_{\processTreeOp}(\ptWfNet,\systemNet)$, and, let $\ptWfNet'{=}(\tilde{\places}',\tilde{\transitions}',\tilde{\netArcs}',\tilde{\place}_i,\tilde{\place}_o,\labelFuncPTWF'){=}\pnptRedution_{\processTreeOp}(\ptWfNet,\systemNet){\in}\univPtWfNet$. 
$\ptWfNet'$ is sound iff $\ptWfNet$ is sound.
\begin{proof} 
($\Rightarrow$) 
Let $\transition'{\in}\tilde{\transitions}'{\setminus}\tilde{\transitions}$. 
Assume, $\ptWfNet$ is sound, yet, $\ptWfNet'$ is not sound.
By definition of any reduction $\pnptRedution_{\processTreeOp}(\ptWfNet,\systemNet)$, if $\ptWfNet'$ is not safe, then $\ptWfNet$ is not safe.
For any $\transition{\in}\tilde{\transitions}{\cap}\tilde{\transitions}'$, if $\nexists{\marking{\in}\reachable(\ptWfNet',[\place_i])}\left((\ptWfNet',\marking)[\transition\rangle\right)$, then also, $\nexists{\marking{\in}\reachable(\ptWfNet,[\place_i])}\left((\ptWfNet,\marking)[\transition\rangle\right)$.
Similarly, if $\splitatcommas{\nexists{\marking{\in}\reachable(\ptWfNet',[\place_i])} \left((\ptWfNet',\marking)[\transition'\rangle\right)}$, then this is also holds for the transitions in $\systemNet$. 
In case $\exists{\marking{\in}\reachable(\ptWfNet',[\place_i])}$ s.t. $\splitatcommas{\nexists{\sequence{\in}\transitions'^*}\left((\ptWfNet',\marking)\xrightarrowdbl{\sequence}(\ptWfNet',[\place_o]) \right)}$, then, again by definition of the reductions, also $\marking{\in}\reachable(\ptWfNet',[\place_i])$ and\\ $\nexists{\sequence{\in}\transitions'^*}\left((\ptWfNet,\marking)\xrightarrowdbl{\sequence}(\ptWfNet,[\place_o]) \right)$. 
($\Leftarrow$) Symmetrical. $\hfill \square$
\end{proof}
\end{lemma}

\begin{lemma}[Pattern Reduction is Language Preserving in $\ptWfNetUnfolding$]
\label{lemma:reduc_lang}
Let $\processTreeOp{\in}\{\to,\times,\wedge,\circlearrowleft\}$, let $\ptWfNet{=}(\tilde{\places},\tilde{\transitions},\tilde{\netArcs},\tilde{\place}_i,\tilde{\place}_o,\labelFuncPTWF){\in}\univPtWfNet$, let $\systemNet{\in}\univPtSystemNet$, s.t., $\pnptPattern_{\processTreeOp}(\ptWfNet,\systemNet)$, and, let $\ptWfNet'{=}(\tilde{\places}',\tilde{\transitions}',\tilde{\netArcs}',\tilde{\place}_i,\tilde{\place}_o,\labelFuncPTWF'){=}\pnptRedution_{\processTreeOp}(\ptWfNet,\systemNet){\in}\univPtWfNet$.
$\langNetVis(\ptWfNetUnfolding(\ptWfNet)){=}\langNetVis(\ptWfNetUnfolding(\ptWfNet'))$.
\begin{proof}
Let $\transition'{\in}\tilde{\transitions}'{\setminus}\tilde{\transitions}$.
By definition of any ${\pnptPattern_{\processTreeOp}(\ptWfNet,\systemNet)}$, $\langNetVis(\systemNet){=}\langPT\left(\labelFuncPTWF'(\transition')\right)$. 
Furthermore, given $\marking,\marking'{\in}\univMultisets(\tilde{\places})$, s.t., for $\sequence{\in}\langNet(\systemNet)$, $(\ptWfNet,\marking){\xrightarrowdbl{\sequence}}(\ptWfNet,\marking')$, then also, $(\ptWfNet',\marking){\xrightarrow{\transition'}}(\ptWfNet',\marking')$.
Consequently, then also, $(\ptWfNetUnfolding(\ptWfNet),\marking){\xrightarrowdbl{\ptToWfNetTranslatorTransB(\labelFuncPTWF(\sequence)))}}(\ptWfNetUnfolding(\ptWfNet),\marking')$ and  $(\ptWfNetUnfolding(\ptWfNet'),\marking)\xrightarrow{\ptToWfNetTranslatorTransB(\labelFuncPTWF(\transition')))}(\ptWfNetUnfolding(\ptWfNet'),\marking')$.
For any $\transition{\in}\tilde{\transitions}{\cap}\tilde{\transitions}'$, s.t., $(\ptWfNet,\marking)[\transition\rangle$, then also $(\ptWfNet',\marking)[\transition\rangle$.
Hence, inference of $\ptToWfNetTranslatorTransB(\transition)$ with the $\ptToWfNetTranslatorTransB$ values of the transitions of $\systemNet$ and, $\ptToWfNetTranslatorTransB(\labelFuncPTWF(\transition')))$, remains the same.
Since removal of some elements of $\systemNet$, and addition of $\transition'$, are the only difference between $\ptWfNet$ and $\ptWfNet'$, i.e., every other element is both in $\ptWfNet$ and $\ptWfNet'$, we conclude that $\langNetVis(\ptWfNetUnfolding(\ptWfNet)){=}\langNetVis(\ptWfNetUnfolding(\ptWfNet'))$. \ $\hfill \square$


\end{proof}
\end{lemma}

\begin{theorem}[\autoref{alg:reduction} is able to find Language-Equal Process Trees]
\label{th:algo_lang_eq}
Let $\wfNet{=}\left(\places,\transitions,\netArcs,\place_i,\place_o,\labelFunc\right){\in}\univWfNets$ and let $\ptWfNet{=}(\tilde{\places},\tilde{\transitions},\tilde{\netArcs},\tilde{\place}_i,\tilde{\place}_o,\labelFuncPTWF){\in}\univPtWfNet$ be the resulting WF-net of \autoref{alg:reduction} on $\wfNet$.
If $\tilde{\transitions}{=}\{\transitions\}$, and, $\netArcs{=}\left\{\left(\place_i,\transition\right),\left(\transition,\place_o\right) \right\}$, then, $\langPT(\labelFuncPTWF(\transition)){=}\langNetVis(\wfNet)$.
\begin{proof}
Observe that, $\ptWfNet$ is sound.
\autoref{lemma:reduction_sound} implies that if we (continuously) revert the reductions applied by \autoref{alg:reduction}, i.e., corresponding to all intermediate assignments of $\ptWfNet$ in \autoref{alg:reduction} are sound.\footnote{As a corollary of this fact, it follows that $\wfNet$ is sound.}
Observe that, \autoref{lemma:reduc_lang} proves that the language of the unfoldings of all the intermediate WF-nets found is the same as well.
Since the labels of the initial WF-net are all members of $\labels{\cup}\{\silent\}$, their unfolding remains the same.
Hence, we deduce $\langPT(\labelFuncPTWF(\transition)){=}\langNetVis(\wfNet)$. $\hfill \square$
\end{proof}
\end{theorem}

\section{Evaluation}
\label{sec:eval}
In this section, we evaluate the proposed algorithm. 
We briefly present the implementation, after which we discuss the experimental setup and the results.

\subsubsection{Implementation}
An implementation of \autoref{alg:reduction} is available\footnote{\url{https://github.com/s-j-v-zelst/pm4py-source/blob/pn_to_pt/scripts/pn_to_pt.py}}, i.e., built on top of the process mining framework PM4Py~\cite{DBLP:journals/corr/abs-1905-06169}. 
Note that, the size of the patterns identified has no influence on the correctness of the algorithm.
Hence, the implementation searches for \emph{binary patterns}, yielding binary trees.
Such a tree can be further reduced, e.g., ${\to}\left(\processTree_1,{\to}\left(\processTree_2,\to\left(\processTree_3,\tau\right)\right)\right)$ corresponds to ${\to}\left(\processTree_1,\processTree_2,\processTree_3\right)$. 

\subsubsection{Experimental Setup}
\begin{figure}[tb]
    \centering
    \includegraphics[width=0.75\textwidth]{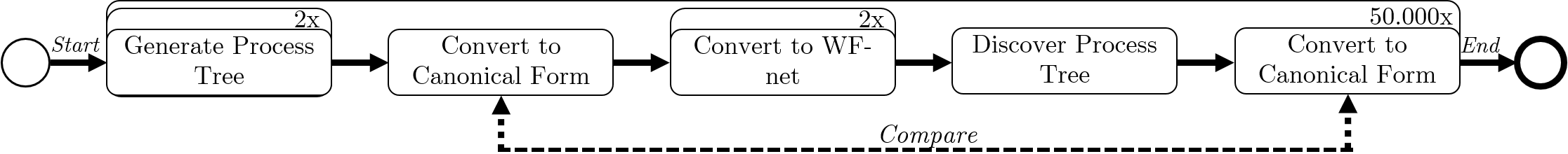}
    \caption{Overview of the experimental setup of the conducted experiments.} 
    \label{fig:exp_setup}
\end{figure}

Here, we briefly discuss the experimental setup of our experiments.
Consider \autoref{fig:exp_setup}, in which we present a graphical overview. 
Using an implementation of \emph{PTandLogGenerator}~\cite{DBLP:conf/bpm/JouckD16,DBLP:journals/bise/JouckD19}, we generate process trees, using two triangular distributions for the number of activities, i.e., $\{10,20,30\}$ and ${\{40,50,60\}}$.
The process trees are translated to WF-nets, using two different translations.
One translation creates invisible \emph{start} and \emph{end} transitions for each operator; the other translation only does so when required (similar to~\autoref{fig:pt_pn_translation}).
The first translation generates larger nets in terms of transitions/places/arcs. 
For each tirangular distribution/translation combination, we generate $50.000$ process trees (yielding $200.000$ experiments).
Finally, we compare the generated process tree in canonical form, to the resulting process tree in canonical form.

\subsubsection{Results}
\begin{wrapfigure}{l}{0.5\textwidth}
\vspace{-45pt}
\centering
    \includegraphics[width=0.5\textwidth]{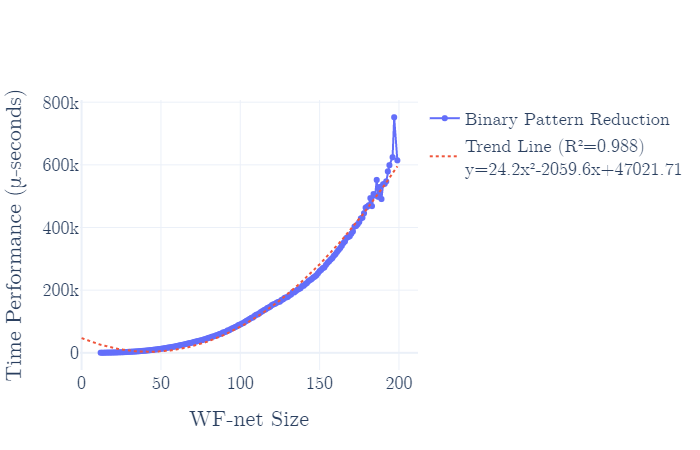}
    \caption{Average time performance of the implementation. A quadratic relation, in computation time ($\mu$-seconds), w.r.t. the size of the WF-net, is observable.}
    \label{fig:res_avg_trend}
\end{wrapfigure}
Here, we briefly discuss the results of the conducted experiments.
Consider \autoref{fig:res_avg_trend}, in which we present the average time performance of the implementation. 
We plot the time performance, conditional to the size of the input WF-net.
Additionally, we plot a polynomial trend-line, computed using polynomial least squares.
As is clearly observable in \autoref{fig:res_avg_trend}, the time performance is quadratic in the size of the net ($|\places|+|\transitions|$).
This is confirmed by the $R^2$-score of the trend-line, i.e,. $\sim0.988$.
In all experiments, the canonical form of the generated process tree equals the canonical form of the (re)discovered process tree.



\section{Related Work}
\label{sec:rel_work}
Process trees are often used in the domain of process mining.
However, a complete overview of the field is outside of scope, hence, we refer to~\cite{DBLP:books/sp/Aalst16}.

Various authors studied translating a certain process modeling formalism to another.
Transformations of \emph{graph-oriented} modeling formalisms, e.g., Petri nets, and \emph{block-oriented} modeling formalisms, e.g., Process Trees, are often studied.
In~\cite{DBLP:journals/ijbpim/MendlingLZ08}, the authors generalize work that transforms (both ways) graph-based process modeling formalisms into \emph{Business Process Execution Language for Webservices (BPEL)} (an XML-based format).
The authors characterize several strategies for such translations.
In this context, the work presented in this paper belongs to the \emph{structure-identification} category.

Of particular interest is the work of van der Aalst and Lassen~\cite{DBLP:journals/infsof/AalstL08,DBLP:conf/otm/LassenA06}, i.e., on translating of WF-nets to BPEL.
In the work, XML fragments of BEPL are generated on the basis of a given WF-net.
The algorithm replaces \emph{components}, i.e., connected, complete subnets with a unique start- and end element. 
Users are additionally allowed to define custom components. 
In~\cite{DBLP:journals/dke/VanhataloVK09,DBLP:conf/wsfm/PolyvyanyyVV10} the notion of the \emph{Refined Process Structure Tree} (RPST) and its computation is introduced.
RPSTs are similar to process trees, i.e., a tree structure describing control-flow behavior.
The works~\cite{DBLP:journals/dke/VanhataloVK09,DBLP:conf/wsfm/PolyvyanyyVV10} focus on computing the RPST of a given BPMN model (the authors indicate that the concepts can be generalized to WF-nets).
However, the discoverd fragments RPST-fragments need to be \emph{canonical}, i.e., they are not allowed to overlap with any other fragment.
Similar to~\cite{DBLP:journals/infsof/AalstL08,DBLP:conf/otm/LassenA06}, the fragments need a single source and a single sink element. 

\section{Conclusion}
\label{sec:conclusion}
In this paper, we presented an algorithm to construct a process tree, on the basis of a Workflow net (WF-net).
The proposed algorithm replaces fragments of the WF-net, that correspond to a process tree operator, i.e., by means of reduction rules.
If the algorithm reduces the WF-net into a net, containing just one transition, there exists a corresponding process tree for the given WF-net, with the same language.
The reduction rules proposed are bidirectionally soundness preserving, hence, in case a process tree is found, the original WF-net is sound.
We have conducted experiments using a prototypical implementation, indicating quadratic time complexity in the net, and, process tree rediscoverability.

\emph{Future Work}
We aim to provide diagnostics w.r.t. the reason why a given WF-net cannot be reduced further, e.g., by assessing if removal of certain elements of the WF-net allows for further reduction.
Alternatively, it is interesting to ``wrap'' certain fragments of the net into an encapsulating transition, after which the search to process tree fragments is continued.
Another interesting direction is the search for structural properties of WF-nets that directly indicate whether a given WF-net corresponds to a process tree.

\newcommand{\SortNoop}[1]{}

\end{document}